\documentclass[conference,10pt]{IEEEtran}
\IEEEoverridecommandlockouts
\usepackage[utf8]{inputenc}
\usepackage{cite}
\usepackage{amsmath,amssymb,amsfonts}
\usepackage{algorithmic}
\usepackage{graphicx}
\usepackage{textcomp}
\usepackage{xcolor}
\usepackage{braket}
\usepackage[ruled,vlined]{algorithm2e}
\usepackage{algorithmic}
\usepackage{subfigure}
\usepackage{url}
\usepackage{bm}
\usepackage{here}
\usepackage{newtxtext}
\usepackage[colorlinks=true,urlcolor=blue,citecolor=blue,linkcolor=black]{hyperref}

\usepackage[normalem]{ulem}
\newcommand{\Del}[1]{}
\newcommand{\Add}[1]{#1}

\usepackage{ulem}
\DeclareRobustCommand{\erase}{\bgroup\markoverwith{\textcolor{red}{\rule[.5ex]{2pt}{0.4pt}}}\ULon}

\begin{document}
\title{Enhancing VQE Convergence for Optimization Problems with Problem-specific Parameterized Quantum Circuits}

\author{\IEEEauthorblockN{Atsushi Matsuo}
\IEEEauthorblockA{\textit{IBM Quantum} \\
\textit{IBM Research -- Tokyo} \\
Tokyo, Japan \\
matsuoa@jp.ibm.com}
\and
\IEEEauthorblockN{Yudai Suzuki}
\IEEEauthorblockA{\textit{Keio University} \\
Kanagawa, Japan \\
yudai.suzuki.sh@gmail.com}
\and
\IEEEauthorblockN{Ikko Hamamura}
\IEEEauthorblockA{\textit{IBM Quantum} \\
\textit{IBM Research -- Tokyo} \\
Tokyo, Japan \\
ikkoham@ibm.com}
\and
\IEEEauthorblockN{Shigeru Yamashita}
\IEEEauthorblockA{\textit{Ritsumeikan University} \\
Shiga, Japan \\
ger@cs.ritsumei.ac.jp}
}

\maketitle

\begin{abstract}
The Variational Quantum Eigensolver (VQE) algorithm is gaining interest for its potential use in near-term quantum devices. In the VQE algorithm, parameterized quantum circuits (PQCs) are employed to prepare quantum states, which are then utilized to compute the expectation value of a given Hamiltonian. Designing efficient PQCs is crucial for improving convergence speed. In this study, we introduce problem-specific PQCs tailored for optimization problems by dynamically generating PQCs that incorporate problem constraints. This approach reduces a search space by focusing on unitary transformations that benefit the VQE algorithm, and accelerate convergence. Our experimental results demonstrate that the convergence speed of our proposed PQCs outperforms state-of-the-art PQCs, highlighting the potential of problem-specific PQCs in optimization problems.
\end{abstract}


\maketitle

\renewcommand{\baselinestretch}{0.96}
\section{Introduction}
\label{sec:intro}
Quantum computing has been attracting significant attention due to its potential for solving complex tasks such as integer factorization \cite{shor}, and \Add{the Boolean satisfiability problem}~\cite{grover}. However, current quantum devices suffer from high error rates and are limited in the size of quantum circuits they can execute \cite{preskillNISQ}, which hinders the execution of quantum circuits for more complicated tasks.

The Variational Quantum Eigensolver (VQE) algorithm has been proposed to utilize these limited quantum devices and has been studied intensively \cite{vqe,vqe-theory,vqe-accelerated, vqe-transitions, vqe-ibm}. The VQE algorithm is designed to find the minimal eigenvalue for a given Hamiltonian. It comprises two parts: one executed on quantum computers and the other on classical computers. The part executed on quantum computers utilizes a shallow quantum circuit with parameters, known as a parameterized quantum circuit (PQC) or a variational ansatz, to prepare a quantum state. By adjusting the parameters, the PQC can generate various quantum states. The expectation value of a given Hamiltonian is then calculated using the created quantum state through sampling outcomes. As the VQE algorithm employs the variational method based on sampling results, designing sophisticated PQCs is crucial for ensuring faster convergence.

The VQE algorithm has potential applications in solving optimization problems by constructing appropriate Hamiltonians \cite{optimization_variational2, cvar, ising}. Despite extensive research in the field, there is limited work on developing PQCs specifically for optimization problems. Existing PQCs face two main challenges: (1) there are only a few known types of PQCs with static structures, and (2) they often yield infeasible solutions for optimization problems.

To overcome these challenges, we propose novel \textit{problem-specific PQCs} for two types of optimization problems. These PQCs are designed to incorporate problem constraints, tailored for each problem. Problem-specific PQCs ensure feasible solutions and significantly reducing search spaces, thereby accelerating the convergence of VQE algorithms.

The remainder of this paper is structured as follows: Section~\ref{sec:background} provides background information on quantum circuits and the VQE algorithm. Section~\ref{sec:proposed} presents the proposed PQCs for two types of optimization problems. In Section~\ref{sec:experiment}, we summarize the experimental results of our proposed PQCs, demonstrating their effectiveness in improving convergence speed. Finally, Section ~\ref{sec:conclution} concludes the paper and highlights potential avenues for future research.

\section{Background}
\label{sec:background}
In this section, we provide an overview of quantum circuits and the VQE algorithm.

\subsection{Quantum Circuits}
Quantum circuits serve as a model for quantum computation \cite{nielsen}, comprising qubits and a sequence of quantum gates. 

Unlike classical bits, which can only be in a state of 0 or 1, qubits in quantum computation can exist in the state $\ket{0}$, $\ket{1}$, or a superposition state. The superposition state is a linear combination of $\ket{0}$ and $\ket{1}$, represented as $\alpha \ket{0} + \beta \ket{1}$, where $\alpha, \beta \in \mathbb{C}$ and $|\alpha|^2 + |\beta|^2 = 1$. The coefficients $\alpha$ and $\beta$ are called amplitudes of the corresponding basis states. An $n$-qubit state can be denoted as $\ket{\psi} = \sum_{k \in {0,1}^n} \alpha_k \ket{k}$, where $\alpha_k \in \mathbb{C}$ and $\sum_{k \in {0,1}^n} |\alpha_k|^2 = 1$. This state can also be represented by a $2^n$-dimensional state vector, such as $(\alpha_0, \alpha_1,...,\alpha_{2^n -1})^T$.

Quantum gates perform specific unitary operations on qubits. They represent which unitary operator is applied to the corresponding qubits. The details of quantum gates used in the proposed PQCs are discussed in Sec.~\ref{sec:proposed}.

\subsection{The VQE Algorithm}
The VQE algorithm aims to find the minimal eigenvalue and corresponding eigenvector of a given Hamiltonian. It leverages the variational principle, which is represented as $\lambda_{\textit{min}} \leq \braket{\psi|H|\psi}$. In this inequation, $H$ represents the given Hamiltonian, $\ket{\psi}$ denotes a quantum state, and $\lambda_{\textit{min}}$ symbolizes the minimal eigenvalue of $H$. According to the variational principle, the expectation value $\braket{\psi |H|\psi}$ for any quantum state $\ket{\psi}$ is greater than or equal to the minimal eigenvalue of $H$.

The VQE algorithm consists of two parts, executed on quantum and classical computers. The quantum computer component involves a shallow quantum circuit with parameters, called a PQC. A PQC prepares a quantum state from an initial state and can generate various quantum states by adjusting its parameters. The expectation values of each term in the given Hamiltonian are obtained by sampling outcomes from the prepared quantum state.

Subsequently, the classical computer calculates the total expectation value by summing the expectation value of each term. It then determines the next set of PQC parameters using classical optimization algorithms, such as the Nelder–Mead algorithm \cite{nelder-mead}, the COBYLA algorithm \cite{COBYLA}, and others \cite{powell,cg,nakanishi-opt}.
Afterwards, the PQC with the updated parameters are used to estimate the expectation values of individual terms in the Hamiltonian again.
The procedure is repeated until the terminating condition is met; i.e., the minimum eigenvalue is obtained, or the number of iteration reaches the maximum number.

\section{Problem-specific PQCs: A Tailored Approach to Constraint Solving}
\label{sec:proposed}
\subsection{Overview of the Problem-specific PQC}\label{sec:overview}
In this subsection, we introduce the general concept of the problem-specific PQC. After mapping each binary variable $x_i$ to qubit $q_i$, we focus on the constraints of an optimization problem. Constraints limit the set of feasible solutions for the optimization problem. We leverage these constraints to dynamically construct a problem-specific PQC that reflects the optimization problem's constraints. This allows us to restrict the unitary transformation provided by the problem-specific PQC while considering the constraints. As a result, we can reduce the set of basis states in the output state vector of the problem-specific PQC to shrink the search space effectively.

For instance, consider an optimization problem with the constraint $\sum_i x_i = 1$. This constraint indicates that exactly one variable must be $1$ while the other variables must be $0$. This type of constraint frequently appears in optimization problems, such as the traveling salesman problem and the job scheduling problem. The constraint $\sum_i x_i = 1$ restricts the set of feasible solutions to the set of basis states corresponding to a W state. A W state is a superposition of states where exactly one qubit is $\ket{1}$ while the other qubits are $\ket{0}$, with equal amplitudes. An $n$-qubit W state is represented as $\ket{W} = \frac{1}{\sqrt{n}} (\ket{10...0} + \ket{01...0} + \ket{00...1})$.

The fundamental concept of the problem-specific PQC is as follows. Let $\mathbb{S}_{\textit{all}}$ be the set of all basis states of $n$ qubits, with $|\mathbb{S}_{\textit{all}}|$ equal to $2^n$. Let $\mathbb{S}_{\textit{feasible}}$ be the set of basis states corresponding to feasible solutions of an optimization problem after mapping variables to qubits. By definition, $\mathbb{S}_{\textit{all}}$ includes $\mathbb{S}_{\textit{feasible}}$. For example, if one feasible solution is $x_0 = 1, x_1 = 0$, and $x_2 = 0$, the corresponding basis state is $\ket{q_0 q_1 q_2} = \ket{100}$. Thus, $\ket{100}$ belongs to $\mathbb{S}_\textit{feasible}$. With the problem-specific PQC, we consider a set $\mathbb{S}_\textit{proposed}$ that includes $\mathbb{S}_\textit{feasible}$ but has a smaller size than $|\mathbb{S}_{\textit{all}}|$. The relation between each set is described as $\mathbb{S}_{\textit{feasible}} \subseteq \mathbb{S}_\textit{proposed} \subseteq \mathbb{S}_{\textit{all}}$.

Using $\mathbb{S}_\textit{proposed}$, the basic concept of the problem-specific PQC is expressed in Eq.~\eqref{eq:outline}.
\begin{equation}
\label{eq:outline}
U_{\textit{proposed}}\ket{0} = \sum_i \alpha_i \ket{\psi_i}, \;\;\ket{\psi_i} \in \mathbb{S}_\textit{proposed}
\end{equation}
$U_{\textit{proposed}}$ denotes a unitary transformation provided by a problem-specific PQC and $\ket{0}$ represents a basis state with all zero indices, which we use as the initial state for the problem-specific PQC. 
Also, $\alpha_i$ indicates the amplitude of $\ket{\psi_i}$. These amplitudes are controlled by parameters of the problem-specific PQC. 
Using an appropriate problem-specific PQC, we can modify only $\alpha_i$ while ensuring that the amplitudes of all other states outside $\mathbb{S}_\textit{proposed}$ are set to $0$. We will explain how the problem-specific PQC functions using examples later.

In many cases, an optimization problem has multiple constraints. For such situations, we create multiple problem-specific parameterized quantum sub-circuits, each of which reflects a corresponding constraint. By properly combining these sub-circuits, we can create a problem-specific PQC that reduces the search space even when the optimization problem has multiple constraints. This approach allows us to explore the feasible solution space more efficiently and potentially find optimal solutions faster.

\Add{Quinones et al.~\cite{quinones2020modeling} and Perelshtein et al.~\cite{ perelshtein2023nisq} consider constructions for different types of constraints than those proposed in this paper. Although our approach differs in terms of the specific constraints and optimization problems considered, we also propose further ideas to improve problem-specific PQCs by restricting the achievable quantum state space and reducing cost for noisy quantum devices.
}

\subsection{Problem-specific PQCs for the TSP}
In this subsection, we introduce problem-specific PQCs tailored for the traveling salesman problem (TSP). The TSP is a well-known NP-hard problem that appears in the field of combinatorial optimization. The problem involves a traveling salesman visiting a set of cities with the objective of finding the shortest possible route that allows the salesman to visit each city exactly once and return to the starting point.

Given an undirected graph $G=(V, E)$, where each edge $(u,v) \in E$ has an associated weight $W_{u,v}$, the goal is to find the Hamiltonian cycle with the minimum sum of edge weights. Let $N=|V|$ and label the vertices $1,..., N$. For a quadratic program, we use $N^2$ variables $x_{v,p}$, where $v$ represents the vertex and $p$ represents its order in a prospective cycle. The quadratic program of the TSP can be formulated as shown in Eq.~\eqref{eq:tsp}. \Add{Note that $x_{v, N+1}$ is interpreted as $x_{v,1}$ in Eq.~\eqref{eq:tsp}.}

\begin{equation}
\label{eq:tsp}
\begin{aligned}
\mathrm{Minimize} \quad &\sum_{(u,v)\in E} W_{u,v} \sum_{p=1}^N x_{u,p} x_{v, p+1}\\
\mathrm{Subject\;to} \quad &\sum_{v=1}^N x_{v,p} = 1, \;\;\; p=1, ..., N \\
&\sum_{p=1}^N x_{v,p}=1, \;\;\;v=1, ..., N \\
&x_{v,p} \in \{0,1\}
\end{aligned}
\end{equation}

In this paper, we propose four distinct PQCs for the TSP. Each PQC possesses unique characteristics, such as the types of constraints considered, the number of quantum gates used, and the number of parameters required. The details of these PQCs will be elaborated upon in Sec.\ref{sec:tsp_proposed_1}, Sec.\ref{sec:tsp_proposed_2}, Sec.\ref{sec:tsp_proposed_3}, and Sec.\ref{sec:tsp_proposed_4}.

\subsubsection{PQCs Satisfying First-Line Constraints Only}\label{sec:tsp_proposed_1}
In this section, we discuss the first proposed PQC which accounts for only the first line of constraints. Each constraint in Eq.~\eqref{eq:tsp} requires exactly one variable to be $1$ while the others are $0$. As previously explained, this type of constraint narrows down the feasible solutions to the bases of the corresponding W state. With a constraint for each $p=1,...,N$, the total number of constraints represented by the first line, $\sum_{v=1}^N x_{v,p} = 1$, is $N$. By mapping binary variables to qubits and taking the tensor product of the $N$ corresponding W states, we limit the search space to $\bigotimes_{p=1}^N \ket{W_p}$. We can ignore other bases outside of $\bigotimes_{p=1}^N \ket{W_p}$ since they do not satisfy the constraints in the first line. It is worth noting that we do not consider the constraints in the second line. Consequently, some bases in $\mathbb{S}_\textit{proposed}$ may not satisfy these second-line constraints. However, the relationship $\mathbb{S}_{\textit{feasible}} \subseteq \mathbb{S}_\textit{proposed} \subseteq \mathbb{S}_{\textit{all}}$ remains valid, and the search space is reduced.

In order to generate W states for use in the VQE algorithm, we must construct quantum circuits that produce these specialized states. Deterministic methods for creating W states of arbitrary sizes have been discussed in earlier studies~\cite{1606.09290, qiskitco6:online}. Traditional W states have equal amplitudes for each base, but for the VQE algorithm, we need to control the amplitudes of each base with parameters as shown in Eq.~\eqref{eq:pw_state}. These parameters are then optimized by a classical optimizer to find the minimum eigenvalue.

\begin{equation}
    \label{eq:pw_state}
    \begin{split}
    &\ket{W(\bm{\phi})} = \sum_i \alpha_{i(\bm{\phi})} \ket{\psi_i},\\
    &\sum_i |\alpha_{i(\bm{\phi})}|^2 = 1,\;\;  \ket{\psi_i} \in \{\ket{10...0},\ket{01...0},  \ket{00...1}\}
    \end{split}
\end{equation}

In Eq.~\eqref{eq:pw_state}, $\ket{\psi_i}$ denotes one of the bases in the corresponding W state where the $i$-th qubit is $\ket{1}$ and the remaining qubits are $\ket{0}$. The amplitude $\alpha_i$ possesses a set of parameters, $\bm{\phi}$, which can be adjusted. It is important to note that  $\bm{\phi}$ may include multiple parameters such as $\{\theta_1, \theta_2,...\}\in \bm{\phi}$. We refer to $\ket{W(\bm{\phi})}$ in Eq.~\eqref{eq:pw_state} as a parameterized W state. 

In this paper, we utilize several quantum gates to create a quantum circuit. These gates include single-qubit gates ($X$ and $R_y(\theta)$) and multi-qubit gates (\textit{Controlled Z} or \textit{CZ}, \textit{Controlled NOT} or \textit{CNOT}, \textit{Controlled SWAP} or \textit{CSWAP}, and \textit{Toffoli} gate). The behavior of these multi-qubit gates depends on the state of the control qubits. The unitary matrices representing these gates are provided in \cite{nielsen}. The $R_y(\theta)$ gate has a parameter $\theta$, which can be adjusted dynamically, while the other gates have fixed matrix elements. Gate indices indicate which qubit the gate is applied to, and for two-qubit gates, the left index represents the control bit, and the right index represents the target bit.

\begin{figure}[tb]
\centering
\includegraphics[scale=0.4]{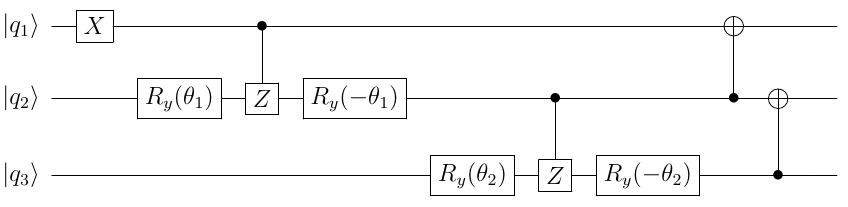}
\caption{Quantum circuit for a parameterized W state of three qubits}
\label{fig:w3_circuit}
\end{figure}

A quantum circuit for a parameterized W state is created using these gates, with an example provided for a case with three qubits, $q_1, q_2$, and $q_3$. The initial state is $\ket{000}$, and a series of gates are applied, including $X$ gates, $R_y(\theta)$ gates, \textit{CZ} gates, and \textit{CNOT} gates. The quantum circuit is illustrated in Figure \ref{fig:w3_circuit}.
The resulting state (Eq. \ref{eq:pw_state_final}) is a parameterized W state of three qubits. 

\begin{equation}
    \label{eq:pw_state_final}
    \begin{split}
    &\alpha_{1(\phi)} \ket{100} + \alpha_{2(\phi)} \ket{010} + \alpha_{3(\phi)} \ket{001}, \;\; \sum_{i=1}^3 |\alpha_{i(\phi)}|^2 = 1,\\
    &\alpha_{1(\phi)} = \cos{\theta_1},\; \alpha_{2(\phi)} = -\sin{\theta_1}\cos{\theta_2},\; \alpha_{3(\phi)} = \sin{\theta_1}\sin{\theta_2}
    \end{split}
\end{equation}

By constructing quantum circuits for parameterized W states, we can develop a problem-specific PQC tailored for the VQE algorithm applied to the TSP. As previously mentioned, a quadratic program representation of the TSP is given by Eq.~\eqref{eq:tsp}. In order to utilize the VQE algorithm, we must map these variables onto qubits. To achieve this, we prepare $N^2$ qubits, denoted as $q_{v,p}$, and associate each variable $x_{v,p}$ with its corresponding qubit $q_{v,p}$. It is important to note that $N$ represents the number of vertices.

\begin{figure}[tb]
	\centering
	\includegraphics[width=0.3\columnwidth]{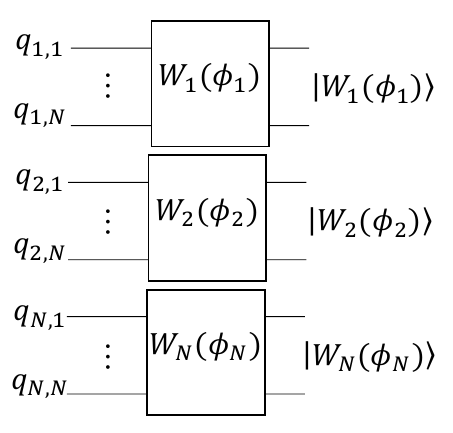}
	\caption{Problem-specific PQC for the TSP, constructed using $N$ distinct quantum circuits to generate parameterized W states, considering only the constraints in the first line of Eq.~\eqref{eq:tsp}}
	\label{fig:tsp_vqe}
\end{figure}

We then employ $N$ distinct quantum circuits to generate parameterized W states with $N$ qubits each. For qubits $q_{1,1}, q_{1,2},..., q_{1,N}$, we introduce the first quantum circuit, which creates an $N$-qubit parameterized W state. Following this, we incorporate the second circuit for qubits $q_{2,1}, q_{2,2},..., q_{2,N}$. We proceed in a similar manner, adding quantum circuits to produce the desired parameterized W states. The final circuit is designated for qubits $q_{N,1}, q_{N,2},...,q_{N,N}$. The resulting quantum circuit is illustrated in Fig~\ref{fig:tsp_vqe}.

In this figure, each box signifies a quantum circuit responsible for creating a parameterized W state with a specific set of parameters $\bm{\phi_i}$ for the associated qubits. Each $\ket{W_i(\bm{\phi_i})} (i=1,...,N)$ on the right side of Fig~\ref{fig:tsp_vqe} represents the output of the corresponding circuit. It is worth noting that the parameter sets for each $\ket{W_i(\bm{\phi_i})} (i=1,...,N)$ are distinct. Utilizing the circuit depicted in Fig.~\ref{fig:tsp_vqe}, we can generate a tensor product of the parameterized W states: $\bigotimes_{p=1}^N \ket{W_p(\bm{\phi_p})}$.

\subsubsection{PQCs Satisfying L-shaped and Remaining Constraints with \textit{CNOT} Gates}\label{sec:tsp_proposed_2}
\begin{figure}[tb]
\centering
\includegraphics[width=\columnwidth]{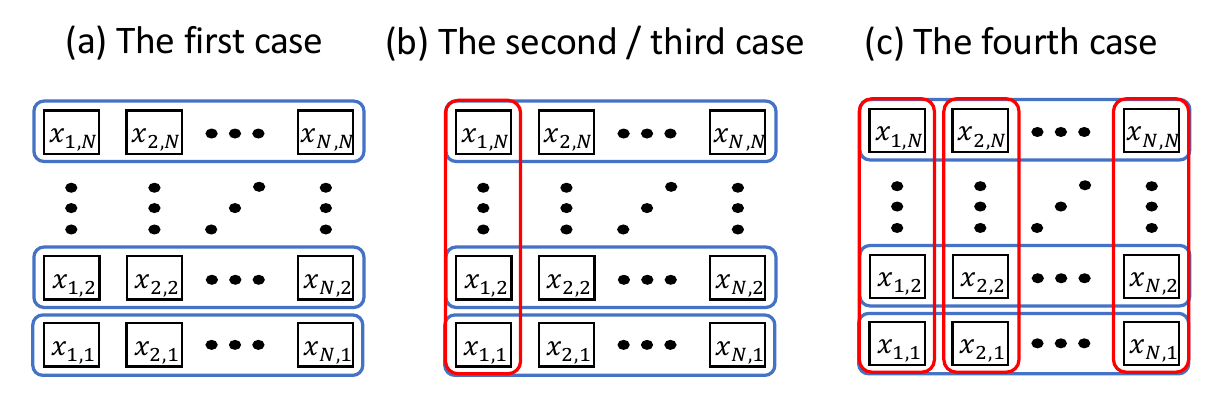}
\caption{Constraints considered for each case: (a) only the first line, (b) an L-shaped constraint constraint and remaining constraints, and (c) all constraints. The blue box represents a constraint from the first line, while the red box indicates a constraint from the second line of Eq.~\eqref{eq:tsp}.}
\label{fig:tsp_constraint_fig}
\end{figure}

In the second PQC, we consider not only the first line of constraints but also a constraint $\sum^{N}_{p=1}x_{1,p} = 1$ in the second line of Eq.~\eqref{eq:tsp}, as illustrated in Fig.\ref{fig:tsp_constraint_fig} (b). This allows for a further reduction of the search space. In contrast to the first PQC, the current scenario demands greater ``correlations'' among qubits mapped from variables, as the variables $x_{1,p}, \; p=1\ldots N$ appear in both the first and second lines. Therefore, it is no longer feasible to realize the constraints using a tensor product of $N$ quantum states. Instead, we employ \textit{CNOT} gates combined with parameterized W state gates to construct the quantum circuit.

The construction of the PQC involves two main steps. In the first step, we create a quantum circuit that satisfies both constraints, $\sum^{N}_{p=1}x_{1,p} = 1$ and $\sum^{N}_{v=1}x_{v,1} = 1$. We refer to this as an ``L-shaped constrain'' due to the L-shape formed by the involved variables in Fig.~\ref{fig:tsp_constraint_fig} (b). In the second step, we apply unitary operations to address the remaining constraints. Detailed descriptions of the construction process and the specific quantum circuits used can be found in Appendix \ref{sec:l-shaped}.

By following this process, we successfully create a problem-specific PQC tailored to address the L-shaped and remaining constraints present in the problem at hand, ultimately leading to a more efficient exploration of the search space.

\subsubsection{PQCs Satisfying L-shaped and Remaining Constraints with Parameter Sharing }\label{sec:tsp_proposed_3}
In the third PQC, we modify the second PQC to reduce the implementation cost for current quantum devices \cite{preskillNISQ}. Current noisy devices suffer from exponential decay of quantum coherence which makes deep circuits problematic. The second PQC becomes deep due to its dependence on constructing a quantum circuit for an L-shaped constraint followed by other gates for the remaining constraints.

To address this issue, we introduce the technique of \textit{parameter sharing}, which results in a shallower circuit with fewer \textit{CNOT} gates. The primary idea behind this technique is to replace \textit{CNOT} gates (and X gates in parameterized W state gates) with $R_{y}$ gates having shared parameters. These parameters ensure that the probability of obtaining specific outcomes is equal to that of other related outcomes. We provide a detailed demonstration of the parameter sharing technique in Appendix \ref{sec:l-shaped-parameter-sharing}.

By leveraging this ``classical correlation", we can create a shallower PQC satisfying the constraints, which is more suitable for current noisy devices. However, the technique has a limitation on its ability to restrict the set of bases compared to the second PQC. 
The quantum state created by the PQC is not fully entangled, i.e. it can be written as a tensor product of smaller quantum states. 
Consequently, the set of bases for the quantum state includes bases that are not present in the second PQC, although the probability of obtaining such additional bases is at most half. This characteristic contributes to interesting results, which we will discuss in Sec.~\ref{sec:experiment}.

\subsubsection{PQCs Satisfying All Constraints}\label{sec:tsp_proposed_4}
We present the fourth PQC, which considers all constraints of Eq. \ref{eq:tsp} to completely exclude infeasible answers as illustrated in Fig.~\ref{fig:tsp_constraint_fig} (c). This ensures that the set of the bases of the quantum states includes only feasible answers, i.e., $\mathbb{S}_{\textit{case 4}} = \mathbb{S}_{\textit{feasible}}$.

The construction of such a PQC for arbitrary $N$ can be achieved in a recursive manner. The basic idea is that the assignments of feasible answers on the 2D grid can be interpreted as permutation matrices, as the constraints of Eq. \eqref{eq:tsp} are exactly the same as the definition of permutation matrices. Utilizing the equivalence between permutation matrices and the assignment of feasible answers on the 2D grid, we can construct the quantum circuit.

We start with a PQC for $N=2$, which can be represented as the superposition of two bases, $\ket{0110}$ and $\ket{1001}$, with the order of qubits $\ket{q_{1,1}q_{2,1}q_{1,2}q_{2,2}}$. Then, we demonstrate that the PQC with $N=k$ can be constructed using the quantum circuit with $N=k-1$ by following a set of procedures. We provide a detailed explanation of these procedures, along with a simple example of the PQC with $N=3$, in Appendix \ref{sec:all-constraints}.

By recursively performing the explained procedure, we can construct the PQC for arbitrary $N$, starting from the quantum circuit with $N=2$. This PQC guarantees that the quantum state is the superposition of bases corresponding to feasible answers, making it suitable for solving the TSP.

\subsubsection{PQCs Satisfying All Constraints with Parameter Sharing}\label{sec:tsp_proposed_5}
In the fifth PQC, we aim to reduce the implementation cost by modifying the fourth PQC, similar to how the third PQC was constructed in Sec.~\ref{sec:tsp_proposed_3}. As described in Sec.~\ref{sec:all-constraints}, the fourth PQC contains many \textit{CSWAP} gates, which are implemented using two \textit{CNOT} gates and one \textit{Toffoli} gate, resulting in a high implementation cost.

To mitigate this issue, we employ parameter sharing as done in the third PQC. In the case of the fifth PQC, we replace \textit{CSWAP} gates with Hop gates (explained in Appendix~\ref{sec:all-constraints-ps}) that have shared parameters, reducing the implementation cost and making the circuit more suitable for current noisy devices. However, like the third PQC, the fifth PQC has a limitation in its ability to restrict the set of bases compared to the fourth PQC due to its reduced entanglement. Further details can be found in Appendix \ref{sec:all-constraints-ps}.

\subsection{Problem-specific PQC for the Minimum Vertex Cover}\label{sec:vertex=cover}
\begin{figure}[tb]
\centering
\includegraphics[width=0.5\columnwidth]{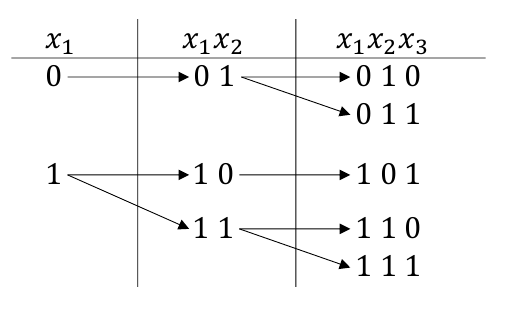}
\caption{Enumerating feasible variable assignments in vertex cover solutions}
\label{fig:condition_vertex_cover}
\end{figure}

In this subsection, we demonstrate the application of the proposed method to the minimum vertex cover problem, another well-known NP-hard combinatorial optimization problem. The minimum vertex cover aims to find the smallest set of vertices in an undirected graph $G=(V,E)$ such that each edge has at least one endpoint in the set. Let $N=|V|$ and label the vertices as $1,\ldots, N$. The problem can be formulated as a linear programming problem as shown in Eq.~\eqref{eq:vertex_cover}, with the number of constraints equal to $|E|$.

\begin{equation}
\label{eq:vertex_cover}
\begin{split}
\mathrm{Minimize} &\sum_{i=1}^N x_i \\
\mathrm{Subject to}\;& (1-x_u)(1-x_v) = 0, \;\; \forall (u,v)\in E \\
&x_{i} \in \{0,1\}
\end{split}
\end{equation}

Each constraint in Eq.~\eqref{eq:vertex_cover} can be interpreted as requiring that if $x_u$ is zero, $x_v$ must be one, and vice versa. To illustrate the process of enumerating feasible variable assignments for the vertex cover, we consider two constraints, $(1-x_1)(1-x_2)=0$ and $(1-x_2)(1-x_3)=0$. As Fig.~\ref{fig:condition_vertex_cover} shows, when $x_1 = 0$, there is only one feasible assignment for $x_1$ and $x_2$ which is $(x_1=0, x_2=1)$. When $x_1 = 1$, there are two feasible assignments for $x_1$ and $x_2$ which are $(x_1=1, x_2=0)$ and $(x_1=1, x_2=1)$.
We then move on to the second constraint $(1-x_2)(1-x_3)=0$. In the same manner, when $x_2 = 0$, there is only one feasible assignment for $x_2$ and $x_3$ which is $(x_2=0, x_3=1)$. When $x_2 = 1$, there are two feasible assignments for $x_2$ and $x_3$, which are $(x_2=1, x_3 =0)$ and $(x_2=1, x_3 = 1)$. By combining the result of the first constraint and that of the second constraint, the feasible assignments of $x_1, x_2$ and $x_3$ can be written as the rightmost column in Fig.~\ref{fig:condition_vertex_cover}.

The construction of a PQC for the minimum vertex cover problem is detailed in Appendix~\ref{sec:pqc_mvc}. By following the approach described in the appendices, we can create a problem-specific PQC that efficiently represents the constraints and structure of the Minimum Vertex Cover problem in order to reduce the search space.

\section{Experimental Results}
\label{sec:experiment}

\begin{table*}[tb]
\centering
\caption{Comparison of required resources for proposed problem-specific PQCs and Ry PQC in the $n$-qubit TSP. Proposed 4 employs \textit{CSWAP} gates, while Proposed 5 employs \textit{Hop} gates instead. $D$ in the Ry PQC corresponds to the depth of the Ry PQC.}
\label{table:tsp_gates_param_comparison}
\footnotesize
\renewcommand{\arraystretch}{0.8}
\begin{tabular}{lccccc}
\hline
& \multicolumn{4}{c}{Necessary Resources} \\
\cline{2-5}
PQCs & Parameters & One-Qubit Gates & Two-Qubit Gates & CSWAP/Hop Gates \\
\hline \hline
Ry PQC & $(\textit{D}+1) n$ & $(\textit{D}+1) n$ & $\textit{D} (n-1)$ & --- \\
Proposed 1 & $n-\sqrt{n}$ & $2n-\sqrt{n}$ & $2n-2\sqrt{n}$ & --- \\
Proposed 2 & $n-\sqrt{n}-1$ & $2n-\sqrt{n}-2$ & $2n-\sqrt{n}-3$ & --- \\
Proposed 3 & $n-\sqrt{n}-1$ & $2n-\sqrt{n}-2$ & $2n-2\sqrt{n}-2$ & --- \\
Proposed 4 and 5 & $\frac{1}{2}n-\frac{1}{2}\sqrt{n}$ & $n-1$ & $n-\sqrt{n}+2$ & $\frac{1}{3}n\sqrt{n}-\frac{1}{2}n + \frac{1}{6}\sqrt{n}-1$ \\
\hline
\end{tabular}
\renewcommand{\arraystretch}{1} 
\end{table*}

In this section, we present the simulation experiments conducted to compare the convergence speed of the proposed Problem-specific PQCs and Ry PQCs using Python. An Ry PQC is one of the most advanced PQCs and it alternates a layer of $R_y$ gates and $CZ$ gates\cite{Qiskit}. The number of layers is called the depth. 
Qiskit Optimization was utilized for converting optimization problems into their corresponding Ising Hamiltonians.
\Add{We used exact state vectors in the simulation experiments. Therefore, only small size problems were experimented with. Experimenting with larger size problems is future work.}

\subsection{Experiments for the Traveling Salesman Problem}
For the TSP, the VQE algorithm in Qiskit was run using the statevector simulator. $10$ trials were conducted with different initial parameters for each PQC. The COBYLA algorithm \cite{COBYLA} served as the classical optimizing algorithm of the VQE algorithm for the TSP. The experiments focused on a complete graph with four nodes for the TSP. All the results presented in this work were obtained with a MacBook Pro with Apple M1 Max and 64GB memories.

\begin{figure}[tb]
	\centering
	\includegraphics[width=\columnwidth]{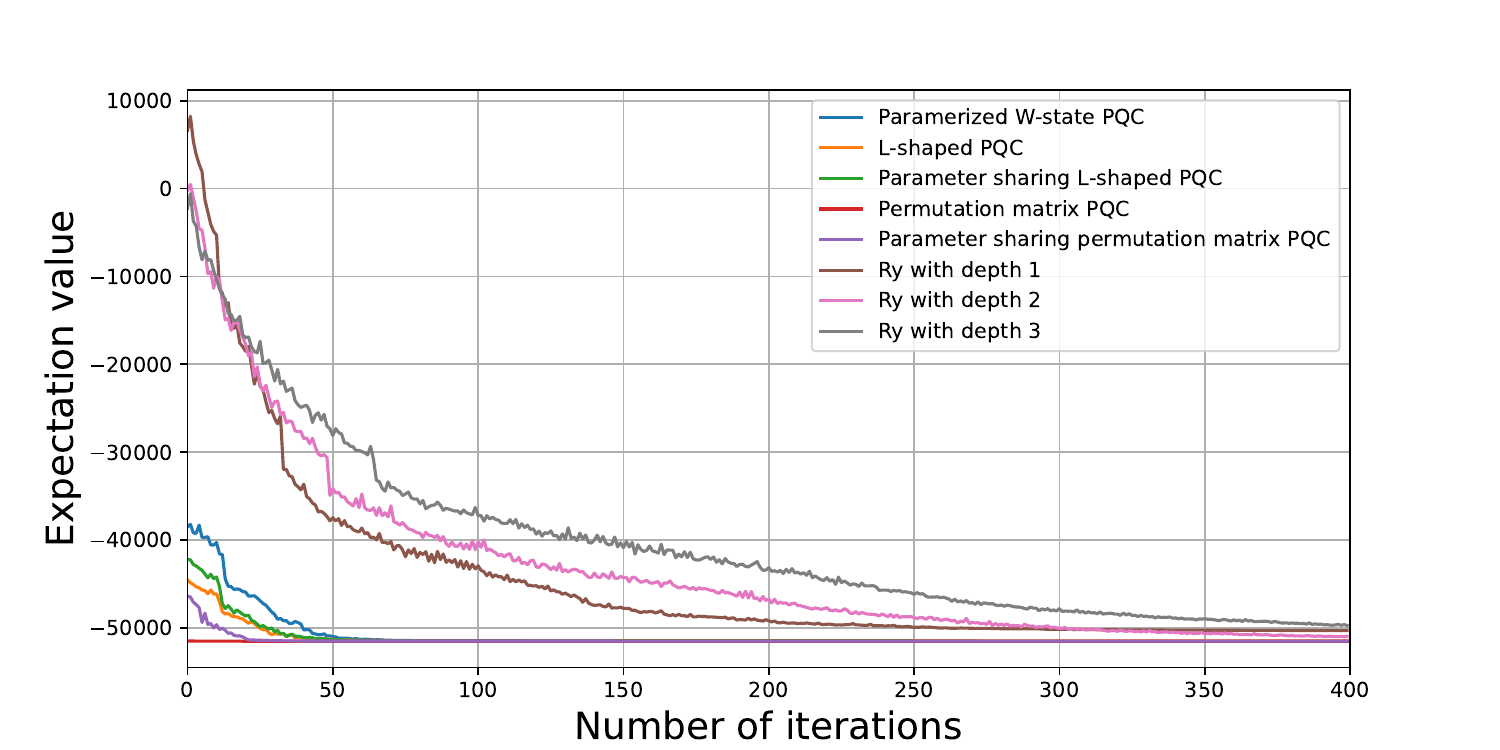}
    \caption{Comparison of proposed problem-specific PQCs and Ry PQCs with depth one, two, and three for the four-city TSP.}
	\label{fig:experiment_tsp}
\end{figure}

\begin{figure}[tb]
	\centering
	\includegraphics[width=\columnwidth]{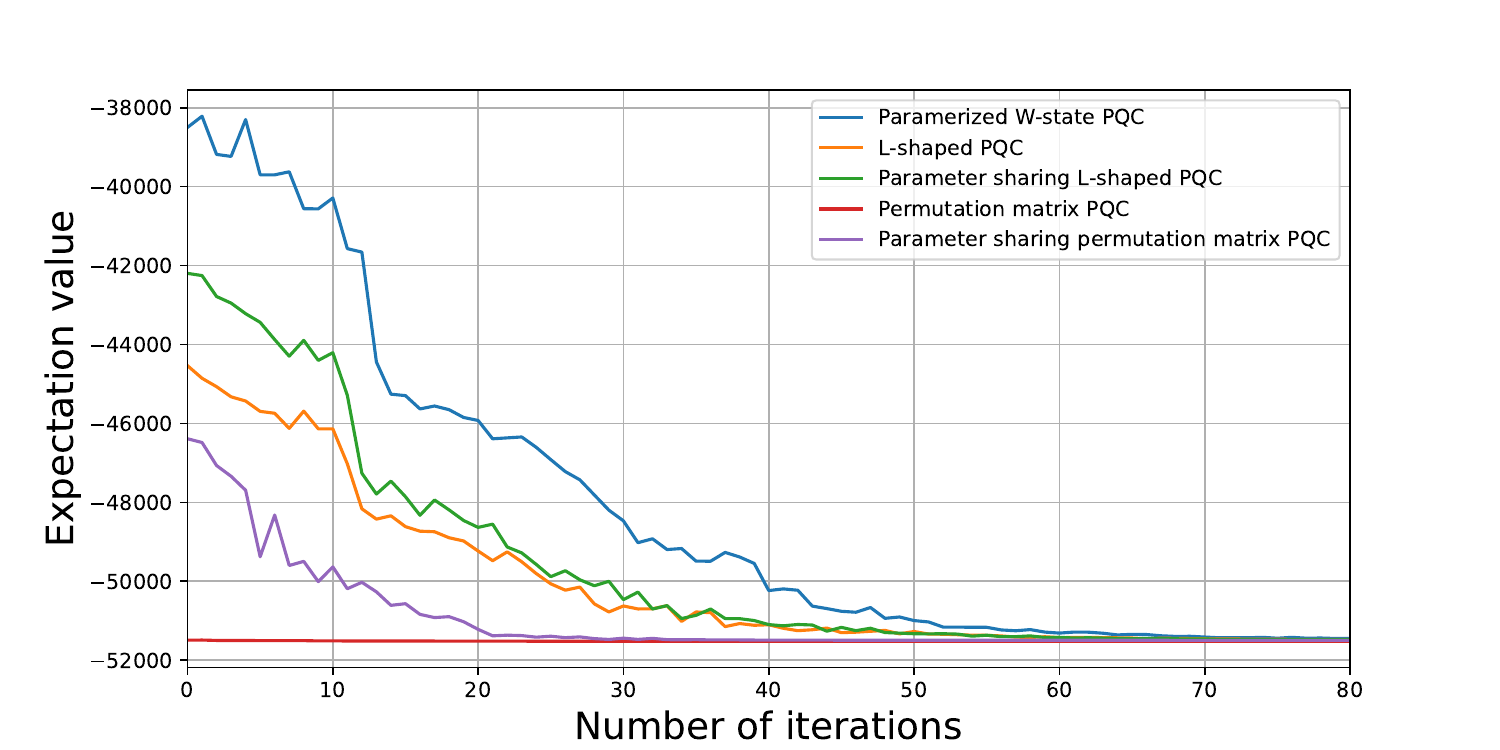}
    \caption{Focused comparison of proposed problem-specific PQCs, extracted from Fig.~\ref{fig:experiment_tsp}.}
	\label{fig:experiment_tsp2}
\end{figure}

Figure~\ref{fig:experiment_tsp} compares the performance of each proposed problem-specific PQC and Ry PQCs with depth one, two, and three. Parameterized W-state PQC (Proposed 1), L-shaped PQC (Proposed 2), Parameter sharing L-shaped PQC (Proposed 3), Permutation matrix PQC (Proposed 4), and Parameter sharing permutation matrix PQC (Proposed 5) correspond to the PQCs detailed in Sec.\ref{sec:tsp_proposed_1}, Sec.\ref{sec:tsp_proposed_2}, Sec.\ref{sec:tsp_proposed_3}, Sec.\ref{sec:tsp_proposed_4}, and Sec.\ref{sec:tsp_proposed_5}, respectively. The proposed PQCs exhibit significantly faster convergence compared to the Ry PQCs. The expectation values of the proposed PQCs decreased rapidly in the first 60 iterations, and their initial expectation values were considerably lower than those of the Ry PQCs. Figure~\ref{fig:experiment_tsp2} presents a graph extracted from Fig.~\ref{fig:experiment_tsp} to focus on the experimental results of the proposed PQCs. 
The convergence time was observed to follow the pattern Proposed 4 $<$ Proposed 5 $<$ Proposed 2 $<$ Proposed 3 $<$ Proposed 1 $<$ Ry, which is closely related to the number of each set of bases.

We also analyzed the ability of each PQC to reach the global minimum. The results show that Proposed 4 reached the global minimum in every trial, while the others did not. Proposed 5 reached the global minimum \Add{nine} times. Proposed 1, 2, and 3 occasionally reached the global minimum, with Proposed 1, 2, and 3 reaching the global minimum three times, four times, and four times, respectively. Ry PQCs did not converge well, producing infeasible answers even after 400 iterations. It is important to note that reaching the global minimum depends on various factors such as problem configurations, types of classical optimizers, and initial parameters.

\Add{The average execution times for the proposed PQCs in the current numerical simulations were $20.22$, $44.41$, $17.5$, $15.5$, and $339.48$ seconds for Proposed 1, 2, 3, 4, and 5, respectively. It is important to note that these values represent the runtime of the numerical simulations until the termination condition is met, and thus, they may not reflect the actual convergence speed of the proposals.
}

Table~\ref{table:tsp_gates_param_comparison} presents the number of necessary gates and parameters for each PQC. From the experimental results, Proposed 4 demonstrated the best performance in terms of convergence speed and reaching the global minimum. However, it requires a significant number of \textit{CSWAP} gates, which are challenging to implement on current noisy devices.

\subsection{Experiments for the Minimum Vertex Cover}

\begin{table}[tb]
\caption{Comparison of required resources for proposed problem-specific PQCs and Ry PQC in the $n$-qubit minimum vertex cover. $D$ in the Ry PQC corresponds to the depth of the Ry PQC.}
\label{table:vc_gates_param_comparison}
\centering
\footnotesize
\renewcommand{\arraystretch}{0.8}
\begin{tabular}{lccc}
\hline
& \multicolumn{2}{c}{Necessary Resources} \\
\cline{2-3}
PQCs & Parameters & Gates \\
\hline \hline
Ry PQC & $(\textit{D}+1) n$ & $(\textit{D}+1) n + \textit{D} (n-1)$ \\
Proposed & $n$ & $3n-2 + n-1$ \\
\hline
\end{tabular}
\renewcommand{\arraystretch}{1}
\end{table}

\begin{figure}[tb]
	\centering
	\includegraphics[width=0.6\columnwidth]{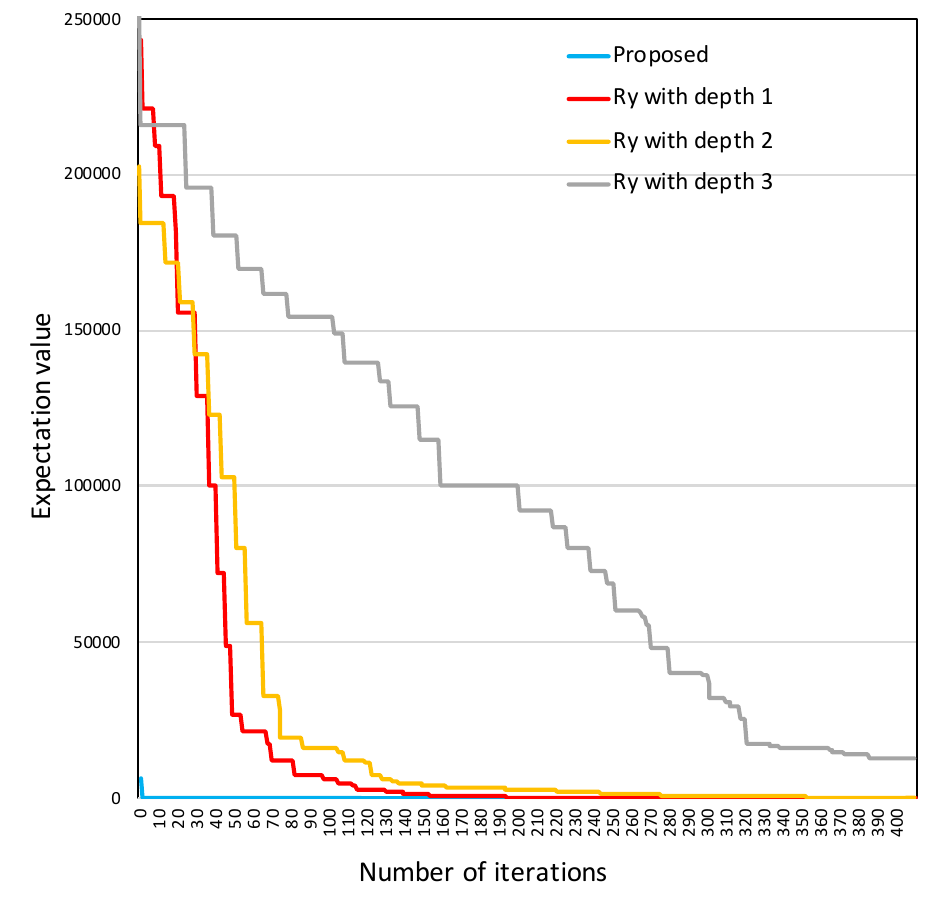}
	\caption{Comparison of the proposed problem-specific PQC and Ry PQCs with depth one, two, and three for the six-node minimum vertex cover.}
	\label{fig:experiment_vc}
\end{figure}

For the minimum vertex cover, NumPy $1.18.4$~\cite{numpy} was used to calculate the expectation values of Ising Hamiltonians, and SciPy $1.4.1$~\cite{scipy} was employed to optimize parameters for the VQE algorithm. The Nelder-Mead algorithm served as a classical optimizer for the minimum vertex cover. The experiments focused on a graph with six nodes containing a cycle. All experiments were conducted on a MacBook Air with a $1.6$ GHz Intel Core i5 processor, DDR3 $8$ GB memory, and macOS $10.14.6$.

Figure~\ref{fig:experiment_vc} compares the performance of the proposed Problem-specific PQC for the minimum vertex cover and Ry PQCs with depth one, two, and three. Similar to the case of the TSP, the proposed PQC exhibited significantly faster convergence than the Ry PQCs. The expectation values of the proposed PQC also decreased rapidly. For example, the expectation value of the proposed PQC after the first iteration was $6342.657$, decreasing to $36.770226$ after the second iteration. The answer for the minimum vertex cover was $3$, and the expectation value of the proposed PQC reached $3.0138958$ after $150$ iterations. In contrast, the expectation value of the Ry PQC with depth 1 reached $1088.1005$ after $150$ iterations and was still $4.729469$ even after $400$ iterations. \Add{The execution time of the current numerical simulation for the proposed PQC was $17.7$ seconds.}

Table~\ref{table:vc_gates_param_comparison} presents a comparison of the necessary parameters and gates between the proposed Problem-specific PQC and the Ry PQCs for the minimum vertex cover. Similar to the case of the TSP, the proposed PQC had a smaller number of parameters than the Ry PQC. Additionally, when the depth of the Ry PQC was large, the numbers of one-qubit gates and two-qubit gates in the proposed PQC were lower than those of the Ry PQC.

Although the amplitudes in the proposed Problem-specific PQCs are not completely independent, they exhibit a slight correlation with each other. This correlation ensures that the amplitudes of the bases corresponding to the optimization problem's answers can be $1$. A careful examination of the relationship between the proposed method for the VQE algorithm and existing methods for classical computers is necessary. Future work will continue to study the convergence to the global minimum and explore strategies for improving the performance of the proposed PQCs.

\section{Conclusion}
\label{sec:conclution}
In this paper, we introduced problem-specific PQCs for the VQE algorithm applied to optimization problems. By focusing on the constraints inherent to a given optimization problem, we were able to dynamically construct PQCs that incorporate these constraints. This approach significantly reduces the search space, leading to faster convergence of the VQE algorithm.

Through numerical experiments, we compared the performance of our proposed PQCs with that of state-of-the-art PQCs. The results demonstrated that our problem-specific PQCs effectively reduced the search space and achieved considerably faster convergence than the current state-of-the-art PQC.

In future work, we aim to explore additional problem-specific PQCs for a wider range of optimization problems and investigate methods to further improve the convergence and implementation efficiency on noisy quantum devices.
\Add{In addition, experimenting with larger size problems is future work. Furthermore, we recognize the need for scalability analysis when dealing with larger problem sizes. 
As the exploration space becomes limited, the energy landscape may become increasingly complex, 
potentially leading to a deterioration in the number of iterations compared to naive strategies. 
In this case, it is crucial to explore strategies to mitigate this issue, which will also be a focus of our future research.
}

\small

\bibliographystyle{IEEEtran}
\bibliography{main}

\newpage
\appendix
\renewcommand{\thefigure}{A.\arabic{figure}}
\setcounter{figure}{0}
\section{Construction of Problem-specific PQCs}
\label{sec:appendices}

\subsection{Construction of PQCs for L-shaped and Remaining Constraints with \textit{CNOT} Gates}\label{sec:l-shaped}
\begin{figure}[tb]
\centering
\includegraphics[width=0.9\columnwidth]{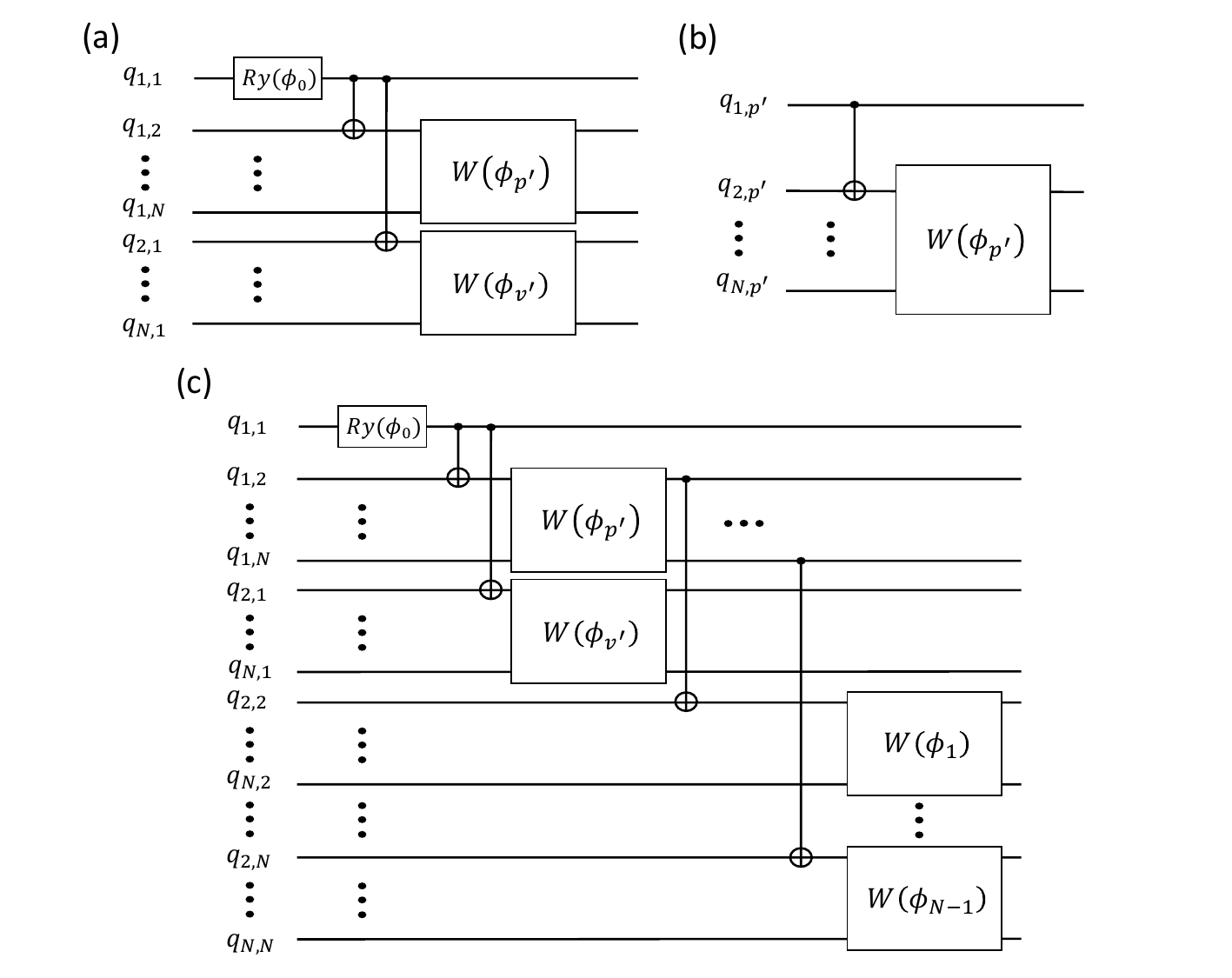}
\caption{PQC Construction for L-shaped and Remaining Constraints with \textit{CNOT} gates: (a) Quantum Circuit for the L-shaped Constraint, (b) Quantum Circuit for the Constraint $\sum^{N}_{v=1}x_{v,p'} = 1$, and (c) Comprehensive View of the PQC}
\label{fig:qc_model2}
\end{figure}

The creation of a problem-specific PQC involves a two-step process. First, we design a quantum circuit that satisfies both the L-shaped constraint, characterized by $\sum^{N}_{p=1}x_{1,p} = 1$ and $\sum^{N}_{v=1}x_{v,1} = 1$, which arises from the interplay of variables in the first and second constraints of Eq.~\eqref{eq:tsp}. Figure~\ref{fig:tsp_constraint_fig} (b) illustrates the L-shape formed by these variables. When $x_{1,1}$ equals zero, one variable from each set $\{x_{1,p'}|p'=2\ldots N\}$ and $\{x_{v',1}|v'=2\ldots N\}$ must equal one, while when $x_{1,1}$ equals one, all variables within these sets must equal zero. Parameterized W state gates can be applied to the qubit sets $\{q_{1,p'}|p'=2\ldots N\}$ and $\{q_{v',1}|v'=2\ldots N\}$ when $\ket{0}_{q{1,1}}$ is present, whereas identity gates are applied when $\ket{1}_{q{1,1}}$ is present. The PQC is then constructed using \textit{CNOT} gates and parameterized W state gates, as shown in Fig.~\ref{fig:qc_model2} (a).

In the second step, we incorporate unitary operations for the remaining constraints, $\sum^{N}_{v=1}x_{v,p'} = 1,\quad p'=2\ldots N$, into the quantum state obtained from the first step. As the qubits corresponding to the variables $\{x_{1,p'}|p'=2\ldots N\}$ in the constraints have already been determined, the constraints can be addressed similarly to the first step. When $x_{1,p'}$ equals one, all variables in the set $\{x_{v',p'}|v'=2\ldots N\}$ are zero, while when $x_{1,p'}$ equals zero, one variable in the set $\{x_{v',p'}|v'=2\ldots N\}$ must equal one. A $CNOT_{q_{1,p'},q_{2,p'}}$ gate, followed by parameterized W state gates on the qubit set $\{q_{v',p'}|v=2\ldots N \}$, can achieve the corresponding unitary operation, as depicted in Fig.~\ref{fig:qc_model2} (b). By employing $N-1$ \textit{CNOT} gates and $N-1$ circuits for parameterized W states, we can create unitary operators that generate a quantum state adhering to the remaining constraints.

By following these two steps, the problem-specific PQC is constructed as shown in Fig.~\ref{fig:qc_model2} (c).

\subsection{Construction of PQCs for L-shaped and Remaining Constraints with Parameter Sharing}\label{sec:l-shaped-parameter-sharing}
We provide a detailed explanation of the parameter sharing technique employed in the construction of the third PQC, which aims to reduce the implementation cost for current quantum devices. This technique results in a shallower circuit with fewer \textit{CNOT} gates while satisfying the constraints. 

The primary idea behind parameter sharing is to replace \textit{CNOT} gates (and \textit{X} gates in parameterized W state gates) used in the second PQC with $R_{y}$ gates that share parameters. These shared parameters ensure that the probability of obtaining specific outcomes is equal to that of other related outcomes, thereby maintaining the desired constraints.

\begin{figure}[tb]
\centering
\includegraphics[width=0.7\columnwidth]{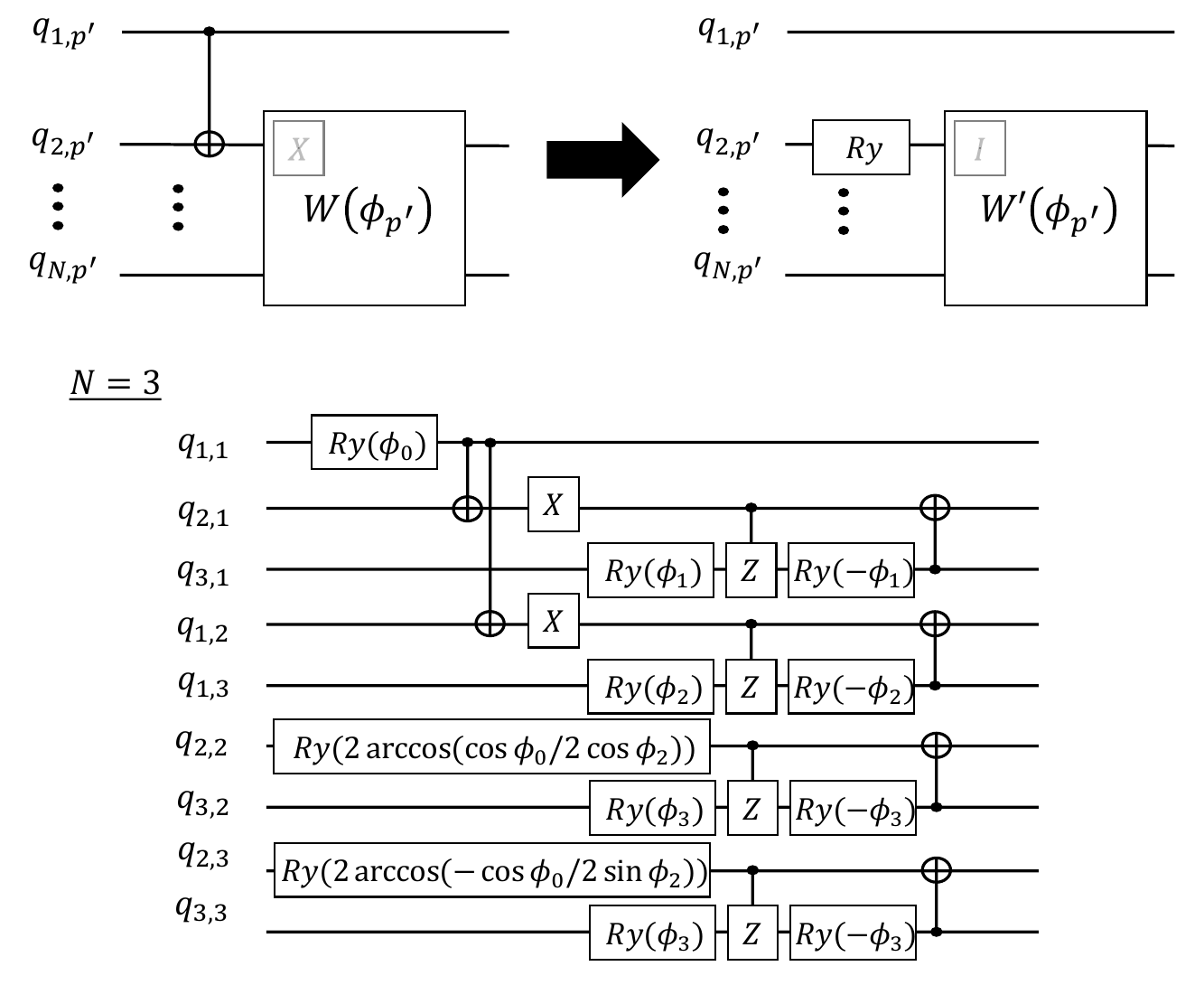}
\caption{Shallower quantum circuit with parameter sharing for the third PQC: Demonstrating an $N=3$ PQC in which the top-left circuit from the second PQC is replaced by the top-right circuit, employing the parameter sharing technique to reduce circuit depth.}
\label{fig:qc_model3}
\end{figure}

To demonstrate the parameter sharing technique in detail, let us consider a simple example of a quantum circuit with $N=3$. As shown in Figure \ref{fig:qc_model3}, parameters $\phi_{0}$ and $\phi_{2}$ are employed for $q_{1,1}$, $q_{1,3}$, $q_{2,2}$, and $q_{2,3}$ in unique ways. These shared parameters are chosen such that the probability of obtaining $\ket{1}_{q_{1,p'}}$ is equal to that of $\ket{0}^{\otimes 2}_{q_{2,p'}, q_{3,p'}}$.
This technique can be easily extended for PQCs with arbitrary $N$ since the probability of obtaining $\ket{1}_{q_{1,p'}}$ for all $p'=2\ldots N$ is analytically calculated in the similar way as shown in Eq.~\eqref{eq:pw_state_final}. By using the shared parameters, the probability of obtaining $\ket{1}_{q_{1,p'}}$ becomes equal to that of $\ket{0}^{\otimes N-1}_{\{q_{v',p'}|v'=2\ldots N\}}$.

\subsection{Construction of PQCs for All Constraints}\label{sec:all-constraints}
We provide a detailed explanation of the construction of quantum circuits for the fourth PQC, which takes into account all the constraints of Eq. \ref{eq:tsp} to exclude infeasible answers. We will present the method for constructing the PQC recursively, starting with an example for $N=2$, and then generalizing the approach to arbitrary $N$.

\ \\
\noindent 
{\bf PQC with $N=2$} 

\begin{figure}[tb]
\centering
\includegraphics[scale=0.35]{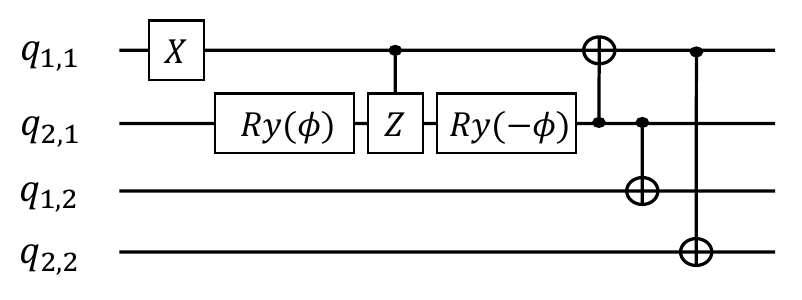}
\caption{Fourth PQC with $N=2$, which creates a superposition of two feasible answers, $\ket{1001}$ and $\ket{0110}$.}
\label{fig:qc_model4_neq2}
\end{figure}

We begin with the construction of the PQC for $N=2$. For this case, we have two feasible answers, represented by two $2\times 2$ permutation matrices,
$\left[
    \begin{array}{cc}
      1 & 0  \\
      0 & 1  \\
    \end{array}
  \right]$ 
and
$\left[
    \begin{array}{cc}
      0 & 1  \\
      1 & 0  \\
    \end{array}
  \right]$. The quantum state we want to create can be described by the superposition of two bases, $\ket{1001}$ and $\ket{0110}$, with the order of qubits $\ket{q_{1,1} q_{2,1} q_{1,2} q_{2,2}}$. A quantum circuit for $N=2$ can be created as shown in Fig.~\ref{fig:qc_model4_neq2}, where the quantum state is represented as $\cos{\phi} \ket{1001} - \sin{\phi} \ket{0110}$.

\ \\
\noindent 
{\bf Recursive Construction for PQCs with $N=k$} 

Next, we describe the recursive construction of PQCs for arbitrary $N=k$. The construction of the PQCs relies on the properties of permutation matrices and their relationship to feasible answer assignments on the 2D grid.

The recursive construction consists of the following steps:

\begin{enumerate}
    \item {\bf Preparation of quantum states for $N=k-1$:} Firstly, we need to prepare the quantum state $\ket{\Psi_{k-1}}$ that represents all feasible answers for $N=k-1$. This quantum state has the order of qubits as $\ket{q_{1,1\ldots} q_{1,k-1}q_{2,1}\ldots q_{2,k-1}\ldots q_{k-1,1}\ldots q_{k-1,k-1}}$. Note that the quantum state for $N=2$ serves as the starting point for the recursive construction.
    
    \item {\bf Initialization of qubits for $N=k$:} Prepare the initialized $2k-1$ qubits, labeled as $q_{k,p'}, p'=2\ldots k$ and $q_{v,k},v=1\ldots k$. These qubits will be used to extend the quantum state $\ket{\Psi_{k-1}}$ to represent the feasible answers for $N=k$.
   
    \item {\bf Application of the parameterized W state gate:} Apply a parameterized W state gate to the set of qubits, $\{q_{v,k}|v=1\ldots k\}$. This step is crucial as it allows us to represent the additional $k$-th row of the $k\times k$ matrix, which can be regarded as the permutation inside the $k$-th row. The parameterized W state gate enables us to manipulate the probabilities of different permutations in the $k$-th row.
    
    \item {\bf Application of \textit{CSWAP} gates:} Apply \textit{CSWAP} gates to the corresponding qubits in $\ket{\Psi_{k-1}}$, $\ket{W_{k}(\psi)}_{\{q_{v,k}|v=1\ldots k\}}$, and $\ket{0}^{\otimes k-1}_{\{q_{k,p'}|p'=1\ldots k-1\}}$. The set of \textit{CSWAP} operations, ${\textit{CSWAP}_{\{q_{v',k},q_{k,p'},q_{v',p'}|p'=1\ldots k-1\} }}$, are applied for all $v'=1\ldots k-1$. The \textit{CSWAP} gates are used to serve as the permutation of the remaining rows depending on the state of the $k$-th row. The states of $\{q_{k,p'}|p'=1\ldots k-1\}$ and $\{q_{v,p'}|p'=1\ldots k-1\}$ are exchanged if $\ket{1}_{q_{v,k}}$, while they remain unchanged if $\ket{0}_{q{p,k}}$.
\end{enumerate}

\begin{figure}[tb]
\centering
\includegraphics[width=0.8\columnwidth]{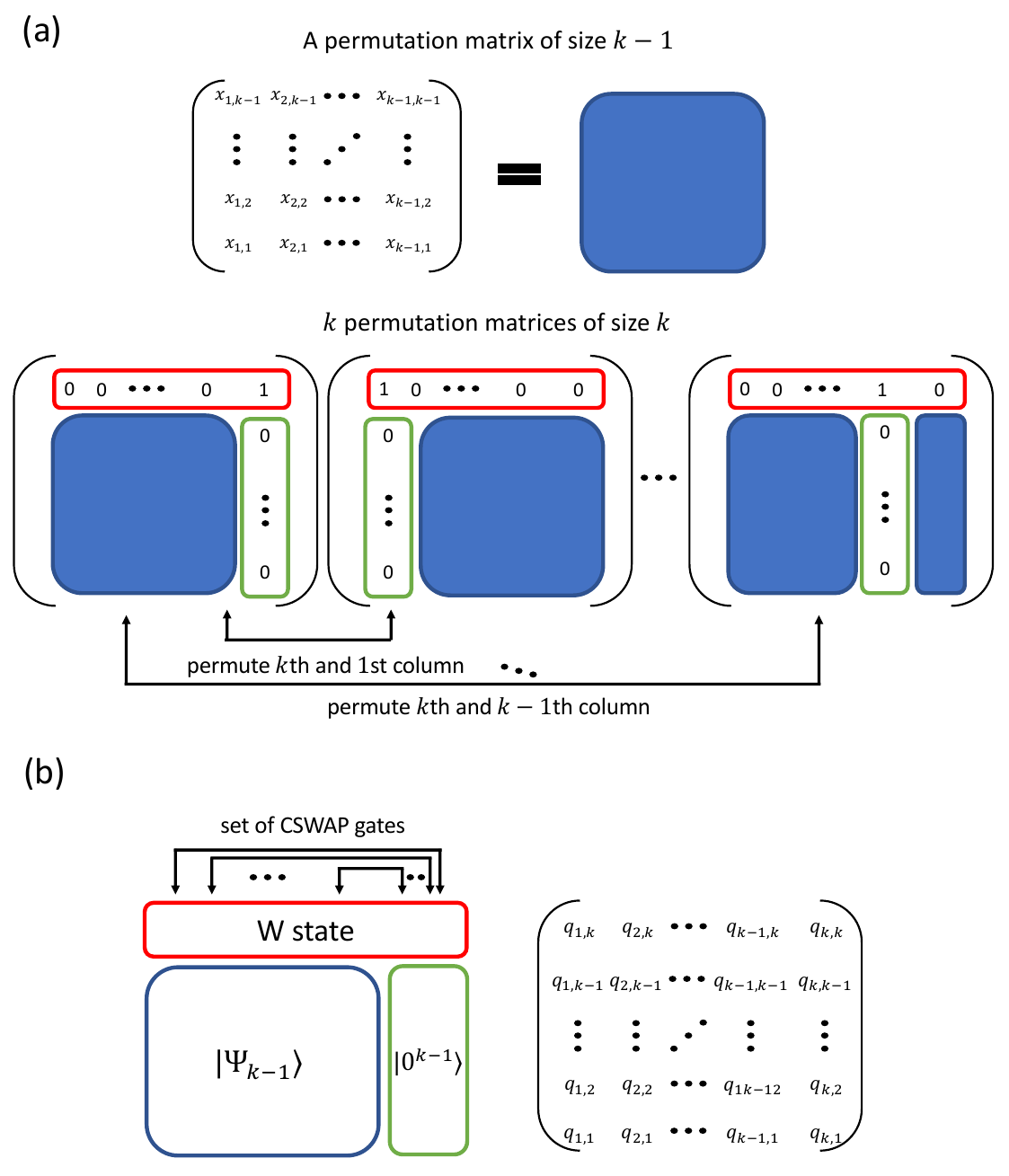}
\caption{Conceptual overview for recursive construction of the forth PQCs: (a) illustrates the property of permutation matrices to obtain permutation matrices of size $k$ using a permutation matrix of size $k-1$ to construct the PQC. (b) shows a schematic view of creating the desired quantum state for $N=k$ using a quantum state for $N=k-1$.}
\label{fig:qc_model4_concept}
\end{figure}

Following these steps, we can construct PQCs for arbitrary $N=k$ that represent the superposition of feasible answers for the traveling salesman problem. The conceptual overview of the recursive construction for the fourth PQC is shown in Fig.~\ref{fig:qc_model4_concept}. The recursive approach allows for efficient construction and manipulation of quantum states for various problem sizes.

\begin{figure}[tb]
	\centering
	\includegraphics[width=0.8\columnwidth]{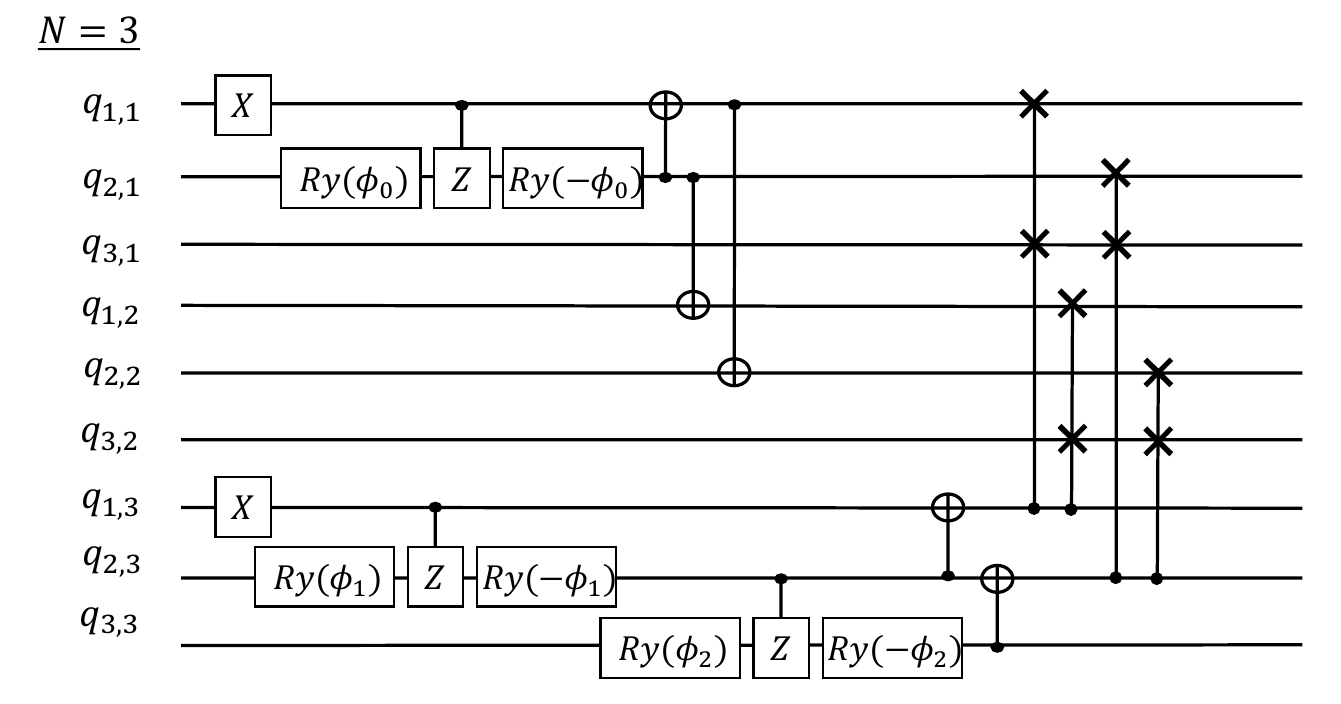}
	\caption{Fourth PQC with $N=3$, which represents the superposition of only feasible solutions for the three-city TSP.}
	\label{fig:qc_model4_neq3}
\end{figure}

As an example, we provide the PQC for the case of $N=3$. The quantum circuit for this case is shown in Fig.~\ref{fig:qc_model4_neq3}. The corresponding quantum state is represented as Eq.~\eqref{eq:state_proposed4_n_3}, with the order of qubits $\ket{q_{1,1}q_{2,1}q_{3,1}q_{1,2}q_{2,2}q_{3,2}q_{1,3}q_{2,3}q_{3,3}}$, which is exactly the superposition of bases of six feasible answers.

\begin{equation}
\label{eq:state_proposed4_n_3}
\begin{split}
\ket{\Psi_{3}}=-\cos{\phi_{0}}\sin{\phi_{1}}\cos{\phi_{2}}\ket{100001010}\\
+\sin{\phi_{0}}\sin{\phi_{1}}\cos{\phi_{2}}\ket{001100010}\\
+\cos{\phi_{0}}\sin{\phi_{1}}\sin{\phi_{2}}\ket{100010001}\\
-\sin{\phi_{0}}\sin{\phi_{1}}\sin{\phi_{2}}\ket{010100001} \\
+\cos{\phi_{0}}\cos{\phi_{1}}\ket{001010100}\\
-\sin{\phi_{0}}\cos{\phi_{1}}\ket{010001100}
\end{split}
\end{equation}

The recursive construction method enables us to represent all possible $N!$ states of feasible answers using $N^2$ qubits. This property is advantageous, as the required memory scales polynomially rather than factorially, which could be a significant improvement over classical computing approaches.

\subsection{Construction of PQCs for All Constraints with Parameter Sharing}\label{sec:all-constraints-ps}
\begin{figure}[tb]
\centering
\includegraphics[width=0.7\columnwidth]{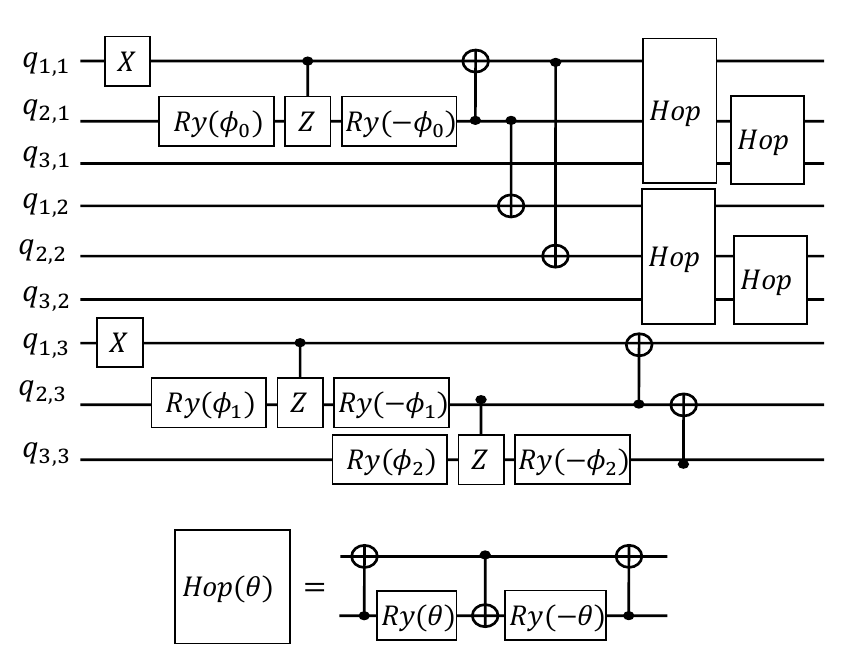}
\caption{Fifth PQC with $N = 3$ by replacing the \textit{CSWAP} gates in Fig.~\ref{fig:qc_model4_neq3} with \textit{Hop} gates. A \textit{Hop} gate can be decomposed into 3 \textit{CNOT} gates and 2 \textit{Ry} gates, as described at the bottom of the figure.}
\label{fig:pqc5}
\end{figure}

We present a detailed construction of the fifth PQC, which utilizes the parameter sharing technique introduced in Sec.\ref{sec:l-shaped-parameter-sharing} and combines it with the fourth PQC discussed in Sec.\ref{sec:all-constraints} to reduce implementation costs. We replace the \textit{CSWAP} gates in the fourth PQC with \textit{Hop} gates, which are represented as $\textit{Hop}(\theta)=\left(\begin{smallmatrix}1 & 0 & 0 & 0 \\ 0 & \sin{\theta} & \cos{\theta} & 0 \\ 0 & \cos{\theta} & -\sin{\theta} & 0 \\ 0 & 0 & 0 & 1\\ \end{smallmatrix}\right)$. A \textit{Hop} gate is a two-qubit gate with a parameter $\theta$, where $\textit{Hop}(0)$ equals a \textit{SWAP} gate. It swaps the states $\ket{01}$ and $\ket{10}$ based on the value of $\theta$. 
We choose the shared parameter in \textit{Hop} gates such that when the control bits of \textit{CSWAP} gates are $1$, the \textit{Hop} gate is equivalent to a \textit{SWAP} gate. Figure~\ref{fig:pqc5} shows the PQC for the case of $N=3$.

\subsection{Construction of Problem-Specific PQCs for the Mnimum Vertex Cover}\label{sec:pqc_mvc}
\begin{figure}[tb]
\centering
\includegraphics[scale=0.5]{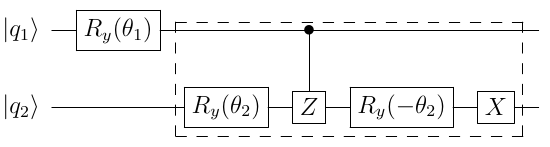}
\caption{Quantum circuit for implementing a single constraint of the Minimum Vertex Cover problem.}
\label{fig:vc_circuit_condition_parameter2}
\end{figure}

In this appendix, we present a method for constructing quantum circuits tailored to the minimum vertex cover problem. By mapping binary variables $x_i$ to qubits $q_i$, we can create a quantum circuit that enumerates feasible variable assignments for the vertex cover, as shown in Fig.~\ref{fig:condition_vertex_cover}. The circuit in Fig.~\ref{fig:vc_circuit_condition_parameter2} demonstrates this process, where the initial state $\ket{q_1 q_2} = \ket{00}$ evolves to the state in Eq.~\eqref{eq:pvc_state} after the circuit is applied. The amplitudes of each basis, $\alpha_{i(\phi)}$, can be controlled by adjusting the parameters $\theta_1$ and $\theta_2$ in the circuit.

\begin{equation}
\label{eq:pvc_state}
\begin{split}
&\alpha_{1(\phi)}\ket{01} + \alpha_{2(\phi)} \ket{10} + \alpha_{3(\phi)} \ket{11},
\; \sum_i |\alpha_{i(\phi)}|^2 = 1,\\
&\alpha_{1(\phi)} = \cos{\frac{\theta_1}{2}}, \; \alpha_{2(\phi)} = -\sin{\frac{\theta_1}{2}} \sin{\theta_2}, \;
\alpha_{3(\phi)} = \sin{\frac{\theta_1}{2}} \cos{\theta_2}
\end{split}
\end{equation}

To accommodate multiple constraints, we can construct a larger quantum circuit by incorporating multiple instances of the sub-circuit shown in the dashed box of Fig.~\ref{fig:vc_circuit_condition_parameter2}. Connecting the sub-circuits requires considering the dependencies among constraints and the topology of the underlying graph. This is crucial because using the sub-circuit in Fig.~\ref{fig:vc_circuit_condition_parameter2} for every constraint when a graph has \textit{cycles} could break the relation $\mathbb{S}_{\textit{feasible}} \subseteq \mathbb{S}_{\textit{proposed}}$. One approach to handle this issue is to construct a spanning tree of the graph and use the sub-circuits only for the constraints corresponding to the edges in the spanning tree. This ensures that the resulting quantum circuit maintains the relationship $\mathbb{S}_{\textit{feasible}} \subseteq \mathbb{S}_{\textit{proposed}}$ and helps reduce the search space. 

\begin{figure}[tbp]
    \centering
    \subfigure[]{%
        \includegraphics[width=0.19\columnwidth]{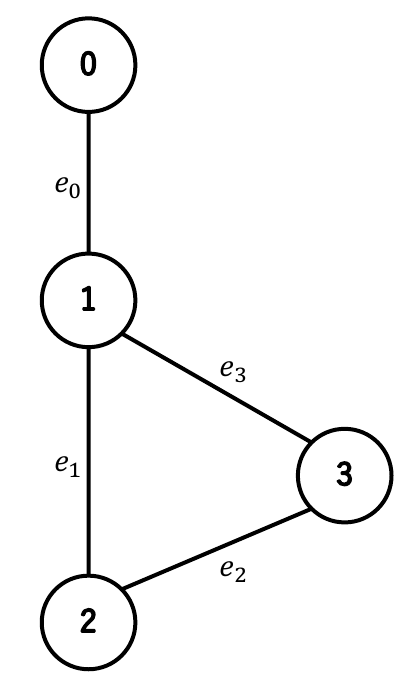}
        \label{fig:vc_4nodeGraph_cycle}}%
    \subfigure[]{%
        \includegraphics[width=0.18\columnwidth]{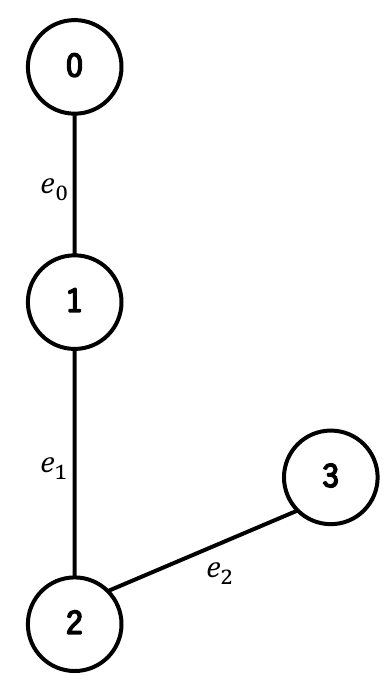}
        \label{fig:vc_spanning_tree}}%
    \subfigure[]{%
        \includegraphics[width=0.6\columnwidth]{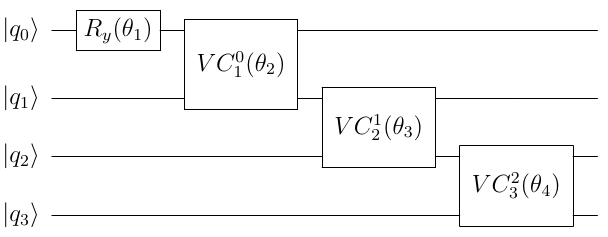}
        \label{fig:vc_circuit_spanning_tree}}%

    \caption{(a) A graph with four nodes and a cycle, (b) a spanning tree of (a), and (c) a problem-specific PQC for the minimum vertex cover based on the spanning tree (b). A sub circuit labeled with \textit{VC} in (c) corresponds to the sub-circuit in the dashed box of Fig.~\ref{fig:vc_circuit_condition_parameter2}.
    }
\label{fig:example_spanning_tree_graph_parameterized quantum}
\end{figure}

For example, consider the graph shown in Fig.~\ref{fig:vc_4nodeGraph_cycle}. A spanning tree of this graph is depicted in Fig.~\ref{fig:vc_spanning_tree}. Figure~\ref{fig:vc_circuit_spanning_tree} illustrates an example of a problem-specific PQC for the minimum vertex cover of the graph based on the spanning tree in Fig.~\ref{fig:vc_spanning_tree}. This approach reduces the number of bases of the search space, yielding a number of bases $< 2^N$ compared to existing circuits, resulting in a more efficient search process.

\end{document}